\documentclass[sigconf]{acmart}
\usepackage{amsmath,url,graphicx,amssymb}
\usepackage{amsfonts}
\usepackage{ctable}
\usepackage{multirow}
\usepackage{algorithm}
\usepackage{algpseudocode}
\usepackage{pifont}
\usepackage{color}
\usepackage{subfigure}
\usepackage{enumitem}
\setcopyright{rightsretained}
\acmDOI{10.475/123_4}

\newcommand{\xeg}{{\it E.g.}}
\newcommand{\eg}{{\it e.g.}}

\newcommand{\etc}{{\it etc.}}
\newcommand{\ie}{{\it i.e.}}

\newcommand{\wrt}{\textit{w.r.t.}}

\ifdefined\setcommenton
\newcommand{\icomment}[2]{{\color{red}$\langle$#1:\textit{\small #2}$\rangle$}} 
\else
\newcommand\icomment[2]{}
\fi

\newcommand{\mylistbegin}{
  \begin{list}{$\bullet$}
   {
     \setlength{\itemsep}{-3pt}
     \setlength{\topsep}{1pt}
     \setlength{\leftmargin}{1em}
     \setlength{\labelwidth}{1em}
     \setlength{\labelsep}{0.3em} } }

\newcommand{\mylistend}{
   \end{list}  }

\newsavebox{\savepar}

\newcommand{\calA}{\mbox{$\mathcal{A}$}}

\newcommand{\calE}{\mbox{$\mathcal{E}$}}

\newcommand{\calG}{\mbox{$\mathcal{G}$}}

\newcommand{\calR}{\mbox{$\mathcal{R}$}}
\newcommand{\calS}{\mbox{$\mathcal{S}$}}

\newcommand{\calV}{\mbox{$\mathcal{V}$}}

\usepackage{euscript}

\newcommand{\deltu}{\mbox{$\mathrel{\hbox{$\bigtriangleup$\raise1.5pt\hbox{\hskip-6.5pt{\tiny $\mu$}}}}$\,}}
\newcommand{\delu}{\mbox{$\mathrel{\hbox{$\bigtriangledown$\raise2pt\hbox{\hskip-6.5pt{\tiny $\mu$}}}}$\,}}

\newcommand{\monus}{\mbox{$\mathrel{\hbox{$-$\raise2pt\hbox{\hskip-5.5pt$\cdot$}}}$\,}}

\newcommand{\semijoin}{\mbox{$\mathrel{\raise1pt\hbox{\vrule height 5pt depth 0pt \hskip-1.5pt$>$\hskip -2.5pt$<$}}$}}

\acmConference[]{}{}{} 
\acmYear{2017}
\copyrightyear{2017}

\acmPrice{15.00}

\begin{document}
\title{Relationship Profiling over Social Networks: Reverse Smoothness from Similarity to Closeness}
\author{Carl Yang}
       \affiliation{
       \institution{University of Illinois, Urbana Champaign}
       \streetaddress{201 N. Goodwin Ave}
       \city{Urbana}
       \state{Illinois}
       \postcode{61801}
       }
       \email{jiyang3@illinois.edu}
\author{Kevin Chen-Chuan Chang}
        \affiliation{
       \institution{University of Illinois, Urbana Champaign}
       \streetaddress{201 N. Goodwin Ave}
       \city{Urbana}
       \state{Illinois}
       \postcode{61801}
       }
       \email{kcchang@illinois.edu}

\begin{abstract}
On social networks, while nodes bear rich attributes, we often lack the `semantics' of why each link is formed-- and thus we are missing the `road signs' to navigate and organize the complex social universe.
How to identify relationship semantics without labels?
Founded on the prevalent \emph{homophily} principle, we propose the novel problem of \emph{Attribute-based Relationship Profiling} (ARP), to profile the closeness \wrt~the underlying relationships (\eg, \textsf{schoolmate}) between users based on their similarity in the corresponding attributes (\eg, \textsf{education}) and, as output, learn a set of \textit{social affinity graphs}, where each link is weighted by its probabilities of carrying the relationships.
As requirements, ARP should be \emph{systematic} and \emph{complete} to profile every link for every relationship--
our challenges lie in effectively modeling homophily:
We propose a novel \textit{reverse smoothness principle} by observing that the similarity-closeness duality of homophily is consistent with the well-known \emph{smoothness assumption} in graph-based semi-supervised learning-- only the direction of inference is reversed.
To realize smoothness over noisy social graphs, we further propose a novel holistic closeness modeling approach to capture `high-order' smoothness by extending closeness from edges to paths.
Extensive experiments on three real-world datasets demonstrate the efficacy of ARP.
\end{abstract}


\maketitle
\section{Introduction}
\label{sec:intro}

While our social universe-- like our social lives-- is complex, they are critically missing `road signs' to navigate. 
On general networks like Twitter and DBLP, the edges (\ie~links, connections) between nodes (\ie~users) are often \emph{unlabeled}-- without `meanings.'
Even on more personal networks like Facebook and LinkedIn-- where we spend much time everday interacting with friends in our ego networks-- our connections with and between friends are lacking the `semantics', in terms of the underlying \emph{relationships}, \eg, \textsf{schoolmate} or \textsf{colleague}, resulting in cluttered social spaces and unorganized interactions.
Such relationship semantics is crucial as `road signs' to organize friends \cite{li2014user,sachan2014spatial,han2015probabilistic} and route information \cite{chakrabarti2014joint,rakesh2016probabilistic,yu2013recommendation} in our social universe. 
Without labeled connections, can we automatically identify the underlying relationships?
This paper aims at such \emph{relationship profiling}, in an \emph{unsupervised} manner, which is important for modeling social networks.

Without pre-defined relationships, what `reasons' do we give as the semantics for each link? 
With the well-known phenomenon of \emph{homophily} \cite{mcpherson2001birds}-- \ie, \emph{the tendency of individuals to stay close with similar others},
it is often the case that a connection between users is a result of such tendency, \ie, it is formed due to their similarity in certain dimensions. 
Moreover, unlike existing works that consider homophily in a single dimension \cite{mislove2010you,csimcsek2008navigating,yang2011like,Yang:2017:BJI:3041021.3054181}, we stress that homophily is naturally \emph{discriminative} in that different relationships correspond to different dimensions of similarity, \ie, different attributes $\calA$ lead to different relationships $\calR$. 

While no social network can capture all possible attributes and relationships, we observe that it is usually trivial to relate the most important relationships in a network to the particular attributes captured there.
\xeg, in a professional network like LinkedIn, the most important relationships are \textsf{schoolmate} and \textsf{colleague}, which are the result of similar \textsf{education} and \textsf{employer} attributes; 
on a personal network like Facebook, friends are formed through relationships such as \textsf{townsmen} and \textsf{hobby peers} resulted from their similarity in \textsf{hometown} and \textsf{hobby}.
Table \ref{table:relationship-types} gives more intuitive examples of important relationships $\calR$ and relating attributes $\calA$ on different networks.

\begin{table*}[ht]
\begin{center}
 \begin{tabular}{|c|c|c|c|c|}
 \toprule
  \hline
 \textbf{Network}&\multicolumn{4}{c|}{\textbf{Important relationships and relating attributes}}\\
 \hline
\multirow{2}{*}{LinkedIn} &$\mathcal{R}$& schoolmate & colleague & professional peer \\
\cline{2-5}
 &$\mathcal{A}$& education background & working experience & skill \\
\hline
\multirow{2}{*}{Facebook} &$\mathcal{R}$ & townsman & hobby peer & acquaintance\\
\cline{2-5}
 &$\mathcal{A}$ & hometown & sports, music, groups, \etc & check-ins, events, \etc  \\
\hline
\multirow{2}{*}{DBLP} &$\mathcal{R}$& research group members & in-field collaborators & cross-field collaborators \\
\cline{2-5}
 &$\mathcal{A}$& publication & paper within the same fields & publication within different fields \\
\hline
  \bottomrule
 \end{tabular}
 \caption{\label{T2}\textbf{Some intuitive examples of relationships $\mathcal{R}$ and attributes $\mathcal{A}$ on different social networks.}}
\label{table:relationship-types}
  \end{center}
 \vspace{-20pt}
\end{table*}

We thus propose the problem of \emph{Attribute-based Relationship Profiling} (ARP), founded on the principle of homophily-- to profile the underlying relationships $\calR$  of each connection by their associated attributes $\calA$.
While the problem is \emph{important}, as social networks strives to help users organize their social universe and route information, it is also \emph{novel}, and we are the first to identify it formally, to the best of our knowledge.

\begin{figure}[t]
\centering
\includegraphics[width=0.48\textwidth]{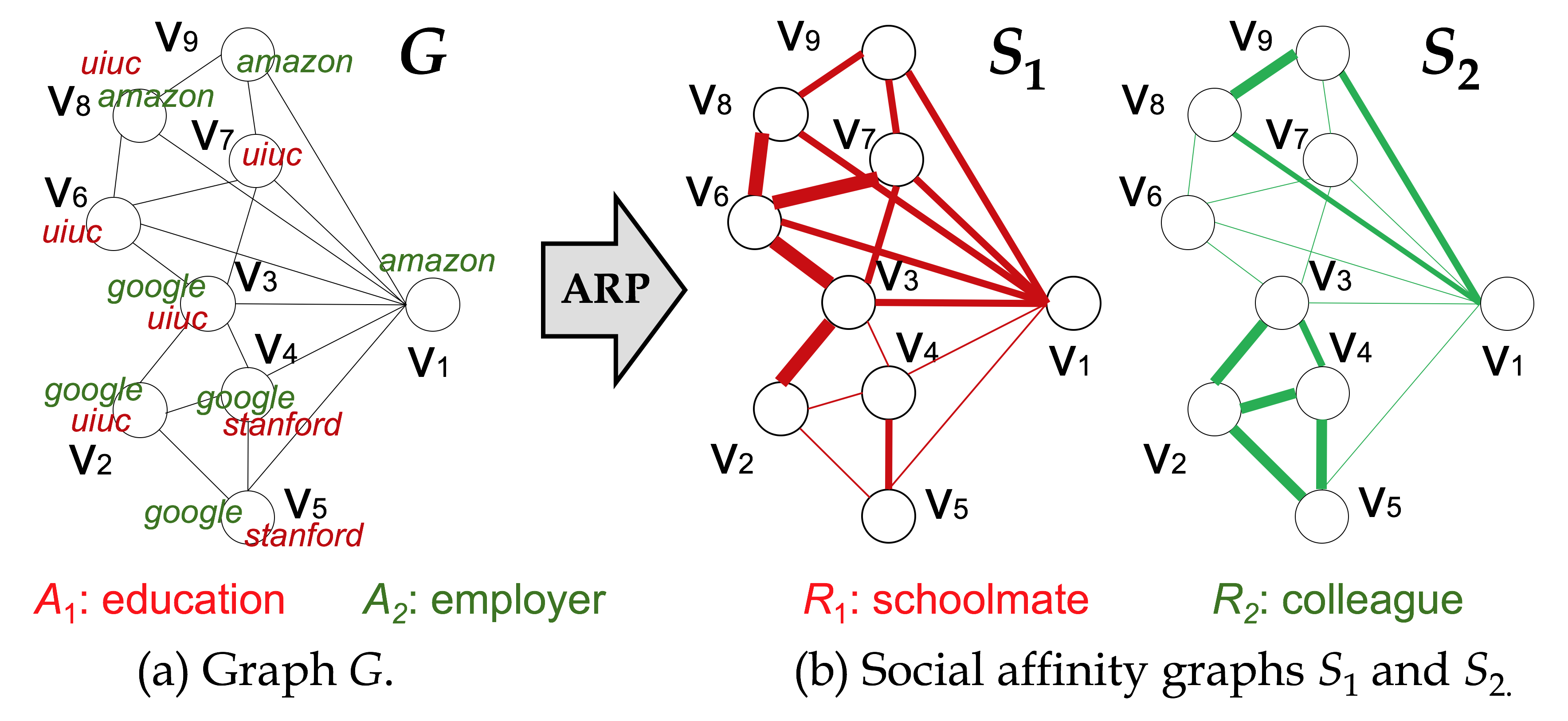}
 \vspace{-20pt}
\caption{A toy example of a simple social network.}
\label{fig:toyexample}
 \vspace{-10pt}
\end{figure}

To illustrate, Figure \ref{fig:toyexample}(a) shows a network $\calG=(\calV, \calE, \calA)$, where $\calV$ is the set of nodes $v_i$ and $\calE$ that of connections $e_{ij}$.
$\calG$ has a set of $M$ important attributes $\calA=\{\calA^m\}_{m=1}^M$ (\eg,  $\calA^1= $ \textsf{education}, $\calA^2= $ \textsf{employer}) with the value functions, \eg, $\calA^1(v_8)= $ \textsf{uiuc}, $\calA^2(v_8)= $ \textsf{amazon}. 
Given $\calG$ as \emph{input},  ARP aims to profile $\calE$ \wrt~$\calA$'s corresponding relationships $\calR=\{\calR^m\}_{m=1}^M$, by inferring its \emph{relationship probabilities} $\{r^m_{ij}=p(e_{ij} |\calR^m, \calA^m)\}_{m=1}^M$, \ie, how each link $e_{ij}$ carries $\calR$.
As \emph{output}, ARP constructs a set of \emph{social affinity graphs} $\calS=\{\calV, \calE, \calR\}=\{\calS^m\}_{m=1}^M$, \ie, graphs sharing the same structure of $\calG$, where each link $e_{ij}$ in $\calS^m$ is now weighted by $r^m_{ij} \in \calR^m$ indicating how it carries $\calR^m$. 
\xeg, as Figure \ref{fig:toyexample}(b) shows, for $\calA^1= $ \textsf{education}, ARP outputs the affinity graph $\calS^1$ for $\calR^1= $ \textsf{schoolmate}, and similarly, for  $\calA^2= $ \textsf{employer}, it outputs $\calS^2$ for $\calR^2= $ \textsf{colleague}. 
To visualize, we plot the thickness of a link to indicate its weight in $\calR$.

We stress that, as the homophily principle implies, ARP should be `systematic' and `complete'. 
On the one hand, individuals stay close because they are similar, and every link should have a probability to carry certain relationships. To this end, our profiling should be \emph{systematic} to cover \emph{every link}. \xeg, two links ($e_{15}$ and $e_{19}$) may not carry a certain relationship (\eg, \textsf{schoolmate}, because it is weak or weaker than other relationships), but we may still want to compare them in that dimension. 
On the other hand, as similar individuals may stay close, more `similarity' leads to more `closeness', and any relationships can co-occur in a link. To this end, our profiling should be \emph{complete} to cover \emph{every relationship} on a link. \xeg, two users ($v_2$ and $v_3$) may be both \textsf{schoolmates} and \text{colleagues}.

While natural, these dual requirements of homophily (and thus relationship semantics) have not been met by most existing works.
Although the problem of ARP is novel, by similarly leveraging the homophily insight, several social mining methods have exploited relationship semantics as their \emph{intermediate results}, but in a rather limited form-- due to the failure to model homophily appropriately (Sec.~\ref{sec:related}):
First, in \emph{attribute profiling} works \cite{chakrabarti2014joint, li2014user,tang2009relational,Yang:2017:BJI:3041021.3054181}, the homophily modeling is \emph{not complete}, by restricting to \emph{one} relationship per link.
Second, in \emph{community detection} works \cite{leskovec2012learning, ruan2013efficient, yang2013community, yang2009combining}, it is \emph{not systematic}, by targeting at each \emph{community} instead of links, which forces links in the same community to carry the same relationship, and leaves out those outside or between communities.
As Sec.~\ref{sec:exp} will show, such improper models fall short for relationship profiling.

Thus, to address ARP, our key challenges center around effectively modeling homophily:

{\flushleft \bf Challenge 1: Systematic and Complete Homophily.} 
As ARP requires, and as the nature of homophily implies, we should realize homophily over every link (systematicness) and for every relationship (completeness), which most existing work failed to satisfy.
What is a principled mechanism for implementing homophily?

{\it \flushleft Insight: Reverse Smoothness Principle.} 
Over a graph, homophily bridges two kinds of `proximities' between users, \ie, \emph{similarity}, measuring how similar two users share for attributes $\calA$, and \emph{closeness}, measuring how close two users link through a relationship $\calR$.
Interestingly, we observe that, as this similarity-closeness duality is natural, it has been explored in graph-based semi-supervised learning (GSSL) \cite{zhou2004learning,zhu2002learning,zhu2003semi}.
GSSL models the \emph{smoothness assumption}, \ie, \emph{points close to each other are likely to share labels}, which helps to infer from closeness (of links) to similarity (of labels) over a given, as \emph{input}, \textit{data affinity graph}, in a systematic (over every link) and complete (for every label) manner.

Surprisingly, while the smoothness assumption is remarkably consistent with homophily, their connection has only been exploited in a non-discriminative way that considers a unique relationship and mixes up all attributes \cite{Yang:2017:BJI:3041021.3054181}. As our key insight is to leverage the modeling of smoothness to realize systematic and complete homophily, we note that our direction of inference for ARP is focused on the opposite to GSSL: from attributes to relationships.
Therefore, we propose the reverse smoothness principle in Sec.~\ref{sec:motiv} and a probabilistic model in Sec.~\ref{sec:model}, to infer from given similarity (of attributes $\calA$) to latent closeness (of relationship $\calR$) and construct, as \emph{output}, a \emph{social affinity graph}. 
We stress that, from similarity to closeness, the focus of ARP is exactly the opposite to that of GSSL-- and this reverse smoothness has not been explored to date. 

{\flushleft \bf Challenge 2: Robust Homophily.} 
While the reverse smoothness principle allows us to relate attributes and relationships by implementing homophily, due to the incompleteness and ambiguity of attributes and links in real-world networks, similarity can not be computed and closeness can not be enforced directly between every pair of nodes.
How to realize homophily robustly over such noisy social networks?

{\em \flushleft Insight: Holistic Closeness Modeling.} Traditional smoothness is only considered on direct edges between pairs of nodes by GSSL on the data affinity graph. 
It works because every edge exists and every pair-wise closeness is enumerated. In real-world networks where attributes are incomplete and ambiguous, and nodes with similar attributes do not always share an edge, such a scheme is useless. 

To deal with real-world networks, we propose a holistic closeness modeling approach in Sec.~\ref{sec:motiv} and implement it in Sec.~\ref{sec:algorithm}, to leverage similarity between every pair of nodes-- even though they may not have a direct link-- by capturing the closeness between nodes based on \emph{paths}, instead of \emph{edges}. In other words, from edges to paths, we extend the traditional smoothness modeling to \emph{higher-order} smoothness, so as to fully exploit the available attribute and link information on an incomplete ambiguous graph. 

We intuitively explain the idea of this approach by continuing on the running example in Figure \ref{fig:toyexample}. It is intuitive to say that $e_{68}$ is very likely to carry relationship \textsf{schoolmate}, because $v_6$ and $v_8$ have the same \textsf{education} attribute. However, tricky questions arise due to the incomplete and ambiguous attributes. \xeg, consider $e_{19}$, where neither of $v_1$ or $v_9$ has available \textsf{education} attribute, and $e_{23}$, where both $v_2$ and $v_3$ have multiple attributes. The holistic closeness modeling approach leverages paths that connect attributed nodes to profile edges they bypass. \xeg, paths $v_8-v_9-v_1-v_7$ and $v_5-v_2-v_3-v_4$ bypass $e_{19}$ and $e_{23}$, respectively, so they add belief on $e_{19}$ to carry relationship \textsf{schoolmate} and $e_{23}$ to carry relationship \textsf{colleague}. In a nutshell, the approach exploits data redundancy in the neighborhood to complete and disambiguate relationships.

{\flushleft \bf Summery.} In this paper, based on our novel reverse smoothness principle and holistic closeness modeling approach, we develop a probability framework of Attribute-based Relationship Profiling (ARP), which leverages user attributes and link structures to reliably estimate the proper relationship semantics in social networks. Specifically, we preserve reverse smoothness on the graph based on an interpretable probability experiment, and we achieve holistic modeling by measuring closeness through standard random walk. An efficient path finding algorithm is designed to solve our justifiable MLE objective. Finally, experiments on three real world datasets comprehensively demonstrate the effectiveness and efficiency of our ARP framework. 
\section{Related Work}
\label{sec:related}
As we discuss in Sec.~\ref{sec:intro}, although we are the first to formally define the problem of ARP, since relationship semantics is critical for various tasks on social networks, algorithms in recent literature have already been intensively solving the related problems to ours. However, while they commonly believe in homophily and connect attributes and relationships with it, they do not correctly interpret the nature of homophily as complete and systematic. According to their main objectives, they can be categorized into two groups. The first group applies homophily to learn attributes through relationships, assuming that relationships on each link are mutually exclusive \cite{chakrabarti2014joint, li2014user, tang2009relational}. While they implicitly learn relationships, they do not compute the complete semantics on each link. The second group utilizes homophily to detect communities, assuming that each community of nodes are connected through the same relationships \cite{leskovec2012learning, ruan2013efficient, yang2013community, yang2009combining}. They compute the semantics of communities, rather than the systematic semantics on links.

The first group of algorithms can produce systematic but not complete relationship semantics.
Since attribute learning aims to infer the missing attributes of every node, systematic relationship semantics can usually be retrieved afterwards by looking at the inferred attributes of nodes on each side of a link.
The recent work EdgeExplain \cite{chakrabarti2014joint} is the closest to ours, which improves on traditional label propagation \cite{zhu2002learning,zhu2003semi} by modeling the interactions among different attributes and optimizing them jointly with relationships. However, it assumes that each link should only carry one relationship.
The discriminative relational learning \cite{tang2009relational} exploits community features as latent social dimensions to aid attribute classification. Therefore, each link is only understood through one attribute chosen by the classification method applied on the two linked nodes.
The Co-Profiling algorithm \cite{li2014user} attempts to learn both user attributes and circles via searching for the reasons of link formation. Each link is then understood through one reason within one of the non-overlapping circles it detects.
The BLA framework \cite{Yang:2017:BJI:3041021.3054181} differentiates links by attribute similarity between the connected nodes, but it does not assume multiple relationships on each link.
Considering two users that are both \textsf{colleagues} and \textsf{schoolmates}, those algorithms force the result to be either of them, which is partial and does not always reflect the truth.
In contrast, ARP will yield two close probabilities \wrt~the two relationships. 

The second group of algorithms can produce complete but not systematic relationship semantics. As they evolve, many community detection algorithms nowadays attempt to characterize communities through attributes. 
Examples include generative models like CESNA \cite{yang2013community} and Circles \cite{leskovec2012learning} and other frameworks like PCL-DC \cite{yang2009combining} and CODICIL \cite{ruan2013efficient}. They all explicitly model the node attributes that cause communities to form and compute a weight matrix characterizing communities \wrt~attributes.
Relationship semantics can then be generated by assuming that links within the same communities carry the same relationships. Therefore, multiple relationships can be associated on each link, if the two linked nodes belong to multiple overlapping communities. 
However, since they only compute the community semantics, the relationship semantics computed from their results are coarse.
To be more specific, there is no way to understand every link, such as those between different communities and outside of any communities. Moreover, they fail to distinguish individual links within the same community. Unlike them, ARP aims to profile relationships in a finer granularity. Rather than relying on the detection of communities, it utilizes the local paths to precisely understand every link as long as a path goes through it.

\section {Motivation}
\label{sec:motiv}
In real-world networks, while links should bear different relationships, they are not explicitly labeled. We argue that being connected in a network does not mean being equally close in reality, and being close does not mean being equally close in every perspective. Since important relationships in social networks are usually discriminatively related with some particular attributes captured by the networks, we propose to leverage user attributes to decipher the hidden relationship semantics of uniform links.

{\flushleft \bf Challenges.} The challenges of ARP, as discussed in Sec.~\ref{sec:intro}, lie in the effective modeling of homophily-- to be systematic and complete as well as robust. The former is difficult due to the lack of a principled way to infer relationships from attributes, and the later is hard because of missing and noisy information in real social networks. 

{\flushleft \bf Principle: Reverse Smoothness.} 
We notice that there is a systematic connection between attributes and relationships as we desire, which has been explored by the principled framework of graph-based semi-supervised learning (GSSL) \cite{zhou2004learning, zhu2003semi}. Specifically, GSSL models two proximities on the graph: \textit{closeness} and \textit{similarity}. Consider GSSL in the social network setting. For each user attribute $\mathcal{A}$, an \textit{affinity graph} $\mathcal{R}$ is used to encode user closeness in terms of the corresponding relationship. Then the value of every user on $\mathcal{A}$ can be learned based on $\mathcal{R}$.

As an example, consider $v_1$, $v_5$ and $v_6$ in Figure 1(a). GSSL assumes that closeness in $\mathcal{R}$ is already given. 
Therefore, if $r_{16}$ is larger than $r_{15}$, the unknown \textsf{education} attribute of $v_1$ will be more likely to be predicted as $a_6$ (\textsf{UIUC}) than $a_5$ (\textsf{Stanford}), due to the following principle of GSSL.
\vspace{-1mm}
\newtheorem{principle}{Principle}
\begin{principle}(Smoothness Principle)
If two nodes $v_i$ and $v_j$ are close on the affinity graph $R$, their attributes $a_i$ and $a_j$ should be similar \cite{zhou2004learning, zhu2003semi}.
\end{principle} 

The focus of GSSL is thus on attribute inference, which goes from closeness to similarity on the graph.

Interestingly, the focus of ARP is the opposite of GSSL, \ie, from similarity to closeness.
In ARP, \eg, we only know there is an edge $e_{13}$ between $v_1$ and $v_3$. We are interested in the closeness on $e_{13}$ in terms of \textsf{schoolmate} and \textsf{colleague}.

Inspired by GSSL, we intuitively reverse the smoothness principle into the following, which allows us to learn $\mathcal{R}$ by systematically enforcing closeness based on similarity, leading to a novel and unique solution to the ARP problem.
\vspace{-1mm}
\begin{principle}(Reverse Smoothness Principle)
If two users $v_i$ and $v_j$ share similar attributes on $\calA$, they should be close on the social affinity graph in terms of ~$\calR$.
\end{principle}

Based on this principle, it is intuitive to implement homophily by probabilistically estimating the closeness on every link in terms of each relationship $\mathcal{R}$ based on the similarity of its related attributes $\mathcal{A}$. The resulting social affinity graphs naturally encode the systematic and complete relationship semantics in the network.

{\flushleft \bf Approach: Holistic Closeness Modeling.}
Our situation in the real-world graph setting is more complex than that of GSSL. While GSSL can enumerate all pair-wise closeness on each edge and enforce similarity accordingly, the opposite is hard to do in social networks with missing and noisy information.

Firstly, attributes are incomplete. Consider $v_1$ and $v_9$ in Figure 1. Since the \textsf{education} attribute $a_1$ and $a_9$ are missing, we have no idea how similar they are, and thus how close $e_{19}$ should imply in terms of \textsf{schoolmate}.
Moreover, even if attributes are complete, closeness cannot be simply enforced on every edge, because similarity can be ambiguous. This is due to the direction of inference, \ie, friends of relationship $\mathcal{R}$ must share the same related attribute $\mathcal{A}$, while similar in $\mathcal{A}$ does not necessarily mean close in $\mathcal{R}$.
\xeg, consider $v_2$ and $v_3$ in Figure 1. If $v_2$ and $v_3$ are \textsf{schoolmates}, they must share the same \textsf{education} attribute such as \textsf{UIUC}. However, sharing the same \textsf{education} attribute does not necessarily imply the relationship of \textsf{schoolmates}. 
In fact, they may be \textsf{colleagues}, because they also share the same \textsf{employer} attribute of \textsf{Google}, or both. If we simply enforce closeness on $e_{23}$, the results will be ambiguous.


To further leverage our reverse smoothness principle and robustly learn the social affinity graph $\mathcal{S}$, we propose to put smoothness constraints and closeness measures onto the whole graph, rather than limiting them to direct edges. Specifically, we define a \textit{path} to be a sequence of non-repeating edges connecting two nodes and use \textit{reachability} to measure closeness as a sum of all weighted paths between two nodes. Then we constrain closeness measured by reachability according to attribute similarity. The intuition is that, the more similar attributes $v_i$ and $v_j$ share, the closer they should be on the graph, and therefore the more paths of shorter lengths and larger weights should connect them.

Continue our example in Figure 1. Inferring $r_{19}$ in terms of \textsf{schoolmate} is challenging due to missing \textit{education} attributes of $v_1$ and $v_9$. However, similarity between $v_7$ and $v_8$ can be used to estimate the closeness on path $v_7-v_1-v_9-v_8$, which indirectly estimates the closeness on $e_{19}$. As a result, $e_{19}$ is likely to carry relationship \textsf{schoolmate}, basically because $v_1$ and $v_9$ share many friends from \textsf{UIUC} such as $v_7$ and $v_8$.

Moreover, continue the discussion from Figure 1 about $v_2$ and $v_3$, where $e_{23}$ is ambiguous. If we combine closeness measured by multiple paths containing $e_{23}$, we will end up with a higher probability of $e_{23}$ to be formed due to \textsf{Google} rather than \textsf{UIUC}, mainly because of the short path $v_4-v_3-v_2-v_5$ containing $e_{23}$ between $v_4$ and $v_5$ with \textsf{Google}. 

By constraining closeness measured by reachability on paths, we effectively utilize the constraints between each pair of nodes $v_i$ and $v_j$ with meaningful attributes onto all edges along the paths connecting $v_i$ and $v_j$, much beyond their direct edges, if any. Among those edges, many are likely to connect nodes without meaningful values of particular attributes, but in this way, they can still get properly constrained and thus well estimated. Moreover, since each edge can be a component of multiple constrained paths, multiple signals from nearby nodes are combined to disambiguate the semantics of that single edge, yielding more robust results.

\section {Model}
\label{sec:model}
Maximizing the production of a similarity term and a closeness term is a standard way of preserving smoothness on the graph \cite{zhou2004learning,zhu2003semi}. However, the objective function is rather heuristically designed for optimization purposes and the scales of learned quantities are arbitrary. 

The objective of ARP is to estimate a complete set of relationship probabilities systematically on each link. Moreover, this has to be done based on incomplete and ambiguous user attributes and link structures. We develop a unified probabilistic framework to derive the objective function of reverse smoothness and precisely learn the proper relationship probabilities through holistic closeness modeling. 

We note that existing probabilistic models in graph-based settings only consider the inference from closeness to similarity on the comprehensive data affinity graphs, instead of the inference in the opposite direction on the incomplete and ambiguous social graphs as we consider \cite{fang2014graph,he2007graph,subramanya2011semi}.


\subsection{Probabilistic Reverse Smoothness}
To learn the systematic and complete relationship probabilities based on user attributes, we apply the reverse smoothness principle by designing a set of simulated probability experiments. 

We start from the description of the probability space. Consider $M$ relationships in a network. We aim to learn one social affinity graph $\calS^m$ by estimating its corresponding relationship probability matrix $R^m$ for each relationship. Since each connection can carry multiple relationships, we assume that $\{R^m\}_{m=1}^M$ follow the multinomial distribution on each existing connection and have $\forall i,j: \sum_{m=1}^M r^m_{ij}=1$.

To estimate $R^m$ based on user attributes, we model the closeness between users $v_i$ and $v_j$ on $\calS^m$, by defining a \textit{user closeness event} that $v_i$ and $v_j$ are close on the graph. We use a random variable $p^m(v_i\sim v_j)$ to denote the probability of this event. 
In a simple case, user closeness can be directly represented by the relationship probability, \ie, $p^m(v_i\sim v_j)=r^m_{ij}$.

Since we only observe attributes $\mathcal{A}^m$ directly from the network rather than the similarity $f^m(a_i,a_j)$, we firstly use intuitive examples to explain how to compute $f^m(a_i,a_j)$.

Consider $\mathcal{R}^m$ = \textsf{schoolmate} and $\mathcal{A}^m =$ \textsf{university}. For a single categorical attribute with multiple distinct values like this, a simple way to compute $f^m(a_i,a_j)$ is to assign 1 to it if $a_i$ and $a_j$ include at least one same value and 0 otherwise. This is basically doing the $or$ operations among the $and$ results between users on each distinct value of the specific attribute. For instance, if $v_1$ has attribute value \textsf{UIUC}, $v_2$ \textsf{Stanford} and $v_3$ both, then $a_1=(1,0)$, $a_2=(0,1)$, $a_3=(1,1)$ and $f^m(a_1,a_2)=0$ while $f^m(a_1,a_3)=f^m(a_2,a_3)=1$.

While we only consider a single categorical attribute towards each relationship in this work, it is trivial to generalize the framework to deal with multiple attributes and numerical attributes, which may help produce better attribute similarities. \xeg, we can consider attributes like \textsf{university} and \textsf{age} for relationship \textsf{schoolmates}. For numerical attributes like \textsf{age}, We can normalize the difference among all users into $[0,1]$ by dividing the largest difference, and similarity then equals one minus the normalized difference. Then we can combine multiple attributes by simply applying a $min(\;)$ function on the similarities. Continue the example above, if user $v_1$ is 50 years old, $v_2$ is 23 and $v_3$ is 20, then we have $a_1=(1,0,50)$, $a_2=(0,1,23)$, $a_3=(1,1,20)$ and $f^m(a_1,a_2)=min(0,0.1)=0$, $f^m(a_1,a_3)=min(1,0)=0$ and $f^m(a_2,a_3)=min(1,0.9)=0.9$. The $min(\;)$ functions make sense because when we consider multiple attributes, we think users are similar only when they are similar in every perspective (e.g., \textsf{schoolmates} should be from the same \textsf{university} and of similar \textsf{ages}). 

With a score $f^m(a_i,a_j)$ computed for each pair of users $v_i$ and $v_j$ describing their similarity on $\mathcal{A}^m$, we estimate their closeness in $\mathcal{R}^m$ in the following simulated probability experiments. 

Each time, we pick up a pair of users $v_i$ and $v_j$ from the sample space $\Omega=\mathcal{V}^2$ according to $f^m(a_i,a_j)$ and observe that they are close on the graph. We require that the probability of randomly picking up $(v_i\sim v_j)$ is proportional to $f^m(a_i, a_j)$. Therefore, considering a total number of $C$ relationships, after a sufficiently large number of experiments, the likelihood of observing the user closeness events is equivalent to
\begin{align}
L= \prod_{m=1}^M \prod_{i=1}^N \prod_{j=1}^N p^m(v_{i}\sim v_j)^{f^m(a_{i},a_{j})}.
\label{EqExp}
\end{align}

By maximizing $L$, we ensure that each pair of users are necessarily close on each social affinity graph $\calS^m$ according to their attribute similarity in $\mathcal{A}^m$, while not too close under the constraints of multinomial distribution. Thus the objective of preserving reverse smoothness over the graph is fulfilled. 

\subsection{Holistic Closeness Modeling}
As we discussed in Sec.~1 and 3, since attributes are incomplete and ambiguous on many users, closeness can not be directly enforced on every connection.
To robustly estimate relationship probabilities, we develop a holistic model of user closeness based on random walks on graphs. While closeness can be asymmetric, we consider it in a symmetric way under the setting of undirected graphs. The framework generalizes trivially to directed graphs.

In this subsection, we derive the closeness model on one social affinity graph $\calS$ and it is exactly the same for all other relationships. Therefore, we use $p$ interchangeably with $p^m$ in this subsection.

In standard random walks, edge weights $\mathcal{R}$ determine the one-step transition probabilities of the random walker on the graph, \ie, $p(v_j|v_i)=\frac{r_{ij}}{d_i}$, where $d_i=\sum_{j\in \mathcal{N}(v_i)}r_{ij}$ and $\mathcal{N}(v)$ is the set of nodes that share an edge with $v$ \cite{fang2014graph,page1999pagerank}. $p(v_j|v_i)$ measures the edge-wise closeness on $\mathcal{S}$.

We propose to further measure path-wise (holistic) closeness.  
Consider a random walk on $\mathcal{S}$. Starting from a specific node $v_i$, besides jumping directly to $v_j$, the random walker can pass through several nodes between $v_i$ and $v_j$ with the corresponding transition probabilities before finally reaching $v_j$. The probability of the random walker to reach $v_j$ from $v_i$ through all possible paths accurately measures the holistic closeness between $v_i$ and $v_j$ on $\mathcal{S}$. 

In order to capture and formalize holistic closeness, we bring out the notion of \textit{reachability} in random walk \cite{lovasz1993random,zhu2013incremental}:
\begin{align}
R(v_i\sim v_j)&=\sum_{l\in l(v_i\sim v_j)}r(l),
\label{EqR0}
\end{align}
where $l(v_i\sim v_j)$ is the set of all paths connecting $v_i$ and $v_j$, and $r(l)$ is the reachability through the specific path $l$ in a random walk. Although we focus on the reachability in only one direction, closeness is modeled symmetrically because we consider similarities in both directions equally.

Suppose all possible paths connecting $v_i$ and $v_j$ are known. We systematically enumerate reachability w.r.t.~paths of different lengths and then add them up into a uniform representation.
Specifically, we use $l_{kh}(v_i\sim v_j)$ to denote the $h$th path of length $k$ between $v_i$ and $v_j$. Suppose $l_{kh}(v_i\sim v_j)=v_{h_1}-v_{h_2}-\ldots- v_{h_{k+1}}$. At each step from $v_{h_i}$ to $v_{h_{i+1}}$, the transition probability is $\frac{r_{h_ih_{i+1}}}{d_{h_i}}$. We also consider the decay factor $\alpha$ to demote the impact of longer walks. Therefore, the reachability under the measure of $l_{kh}(v_i\sim v_j)$ is
\begin{align}
p_{kh}(v_i\sim v_j)=\alpha^k\prod_{s=1}^k\frac{r_{h_sh_{s+1}}}{d_{h_s}}.
\label{EqR1}
\end{align}
In this form of multiplication, since the weight of the whole path is proportional to the weight of each edge and sub-path along that path, the closeness among nodes is naturally coupled and transmitted along the path.

Supposing there are totally $H$ paths of length $k$ connecting $v_i$ and $v_j$, then we have
\begin{align}
p_{k}(v_i\sim v_j)&=\sum_{h=1}^Hp_{kh}.
\label{EqR2}
\end{align}

Considering all paths of different lengths between $v_i$ and $v_j$, we have
\begin{align}
p(v_i\sim v_j)&=\sum_{k=1}^Kp_{k},
\label{EqR3}
\end{align}
where $K$ is the maximum length of paths we consider. To fully implement reachability as in Eq.\ref{EqR0}, $K$ should be set to $+\infty$. However, it is usually sufficient to set $K$ to small numbers like 3 or 4, due to the small world phenomenon, which makes longer paths less important \cite{watts1998collective}. According to \cite{zhu2013incremental}, the ignored reachability on paths longer than K is bounded by $\alpha^{K+1}$, and in practice, we can dynamically increase $K$ to compute incremental reachability. In Sec.~\ref{sec:exp}, we show the impact of different $\alpha$ and $K$.

Combing Eq.\ref{EqR1}, Eq.\ref{EqR2} and Eq.\ref{EqR3}, we get the reachability between $v_i$ and $v_j$ measured by the whole graph as

\begin{align}
p(v_i\sim v_j)&=\sum_{k=1}^K\sum_{h=1}^H\alpha^k\prod_{s=1}^k\frac{r_{h_sh_{s+1}}}{d_{h_s}}.
\label{EqR}
\end{align}

However, finding all possible paths connecting $v_i$ and $v_j$ is non-trivial. Therefore, an efficient path enumerating algorithm is devised especially for our scenario in Sec.~\ref{sec:algorithm}.

\subsection{Interpretation}
We give an interpretation of how our probabilistic framework works in a random-walk perspective.

Combining Eq.\ref{EqExp} and Eq.\ref{EqR}, we get the likelihood function connecting path-wise user closeness with attribute similarity,
\begin{align}
L= \prod_{m=1}^M \prod_{i,j} (\sum_{k=1}^K\sum_{h=1}^H\alpha^k\prod_{s=1}^k\frac{r^m_{h_sh_{s+1}}}{d^m_{h_s}})^{f^m(a_i,a_j)}.
\label{EqLike}
\end{align}

Consider a random walk on the graph.
By constraining edge weights through path-wise closeness measured by reachability on the graph, we are actually requiring the random walker to `prefer' paths connecting nodes with similar attributes, instead of always choosing an edge to go with uniform probabilities. This idea is similar to the supervised random walk in \cite{backstrom2011supervised}. But instead of generating ad hoc features for edges, we directly manipulate edge weights through paths, and thus the actual correspondence between edges and paths is leveraged. 

As a result, for each edge or sub-path, the more paths connecting nodes with similar relating attributes pass through it, the more probable it will be visited by the random walker in the stationary distribution, and thus is more probable to be formed due to the relationship under consideration. 
Moreover, since there are many paths connecting each pair of attributed nodes and each path consists of multiple edges and sub-paths, many relationships among un-attributed nodes can be effectively profiled given only a few pairs of attributed nodes. 
Finally, all considered relationships compete on each connection due to the constraints of multinomial distribution, and the probability of each relationship to be carried on one connection is appropriately related to the number and distance of users with similar relating attributes around it.
Thus the problems of missing and overlapping attributes are systematically addressed.

To show that our model essentially preserves reverse smoothness, we extract the generalized objective function of SSL as
\begin{align}
J_{SSL}=\sum_{m=1}^{M}\sum_{i,j}C^m_{ij}S^m_{ij},
\label{EqGen}
\end{align}
where closeness (C) and similarity (S) are implemented in various ways due to different intuitions and measurements \cite{zhu2002learning,zhou2004learning,lin2014geodesic}. Maximizing Eq.\ref{EqGen} with proper regularization essentially preserves smoothness by reducing the difference between C and S in M dimensions. 

To contrast, we write the log-likelihood of ARP from Eq.\ref{EqLike} as
\begin{align}
J_{ARP}=\sum_{m=1}^{M}\sum_{i,j}f^m(a_i,a_j)\log(\sum_{k=1}^K\sum_{h=1}^H\alpha^k\prod_{s=1}^k\frac{r^m_{h_s h_{s+1}}}{d^m_{h_s}}).
\label{EqARP}
\end{align}
In this equation, $f(a_i,a_j)$ implements S while $\log(p(v_i\sim v_j))$ implements C. The correspondence between Eq.~\ref{EqGen} and \ref{EqARP} indicates the effectiveness of ARP in preserving the reverse smoothness on the social affinity graph.

Note that, unlike Eq.~\ref{EqGen} that is designed purely based on intuitions and optimization purposes, our Eq.~\ref{EqARP} is derived from a principled probabilistic framework, where probability interpretation of relationship semantics is naturally preserved, and the coupling of closeness and similarity is decided by the well defined simulated probability experiments.




\section {Algorithm}
\label{sec:algorithm}

Realizing our holistic smoothness model is to compute a parameter configuration $\mathcal{R}$, so that the likelihood of observing the user closeness event is maximized according to attribute similarity. For this purpose, we need to firstly find relevant paths that can be constructed by existing edges on the graph, and then optimize weights $\mathcal{R}$ on them.

\subsection{Finding Paths on Graph}
According to Eq.~\ref{EqLike}, for each social affinity graph $\calS$, we need to find paths $l(v_i\sim v_j)$ for the pairs of nodes $v_i$ and $v_j$ with $f(a_i,a_j)>0$. Unlike traditional path enumeration on graphs, our problem is quite unique, where we only care about \textit{short} paths between \textit{a small portion} of nodes. 

Since shorter paths contribute more in our model,  we devise an efficient path finding algorithm based on breadth-first search (BFS), so that we can tune path length $K$ to avoid considering longer paths. In practice, $f(a_i,a_j)$ is usually very sparse, since there are numerous distinct values on $\mathcal{A}$ and many users do not have any meaningful value. Therefore, we only need to start from a very small number of nodes compared to $|\mathcal{V}|$. Finally, since we need to record the exact paths and avoid repeated iterations when considering the same nodes in different levels of search, we borrow the path descriptor $d(\cdot)$ from \cite{rubin1978enumerating} to encode, record and retrieve paths between nodes with time complexity $O(1)$. It is also efficient to check if a certain node or edge is on a path and if two paths are the same by simply doing binary AND and XOR on $d$.

Algorithm 1 formally demonstrates our novel path finding method. We evaluate its correctness by checking the completeness and non-repetitiveness. In \textit{Step 8}, by requiring $v_j \notin l$, we require that the next node to be propagated to is not already in the path being considered, so no loopy paths can be generated; by requiring $l+e_{ij} \notin D(I,j)$, we guarantee that each path is generated only once. Moreover, in \textit{Step 10} and \textit{Step 14}, we ensure that the same nodes are not considered multiple times in different search steps. Finally, in \textit{Step 6}, we always consider every possible direction to make sure that no simple path is missed.

Since path indexing and legitimacy checking are efficiently $O(1)$ with the path descriptor lists, the overall computational complexity of finding paths is $O(K|\mathcal{V}|^2)$. However, the actual computational time is much shorter than $K|\mathcal{V}|^2$. In each step of BFS, the numbers of considered nodes and neighbors are much less than $|\mathcal{V}|$. 
The efficiency of finding paths can be further improved by path caching and reusing, to fully utilize the path indexes. Specifically, we try to cache as many legitimate paths as possible after they are indexed. Therefore, the paths of length $K$ can be directly reused when considering paths longer than $K$. As the number of paths goes exponentially with the length of the paths, it is usually impossible to keep all path indexes in cache and even memory. Motivated by the scale-free property of social networks \cite{choromanski2013scale}, which leads to the frequent reuse of paths between a small number of high-degree hub nodes, we adopt the Least Recently Used (LRU) algorithm for path caching. 

\begin{algorithm}[h!]
\caption{Efficient Path Finding Algorithm}
\begin{algorithmic}[1]
\Procedure{EPF}{}
\Comment{Input}

$\mathcal{G}(\mathcal{V}, \mathcal{E})$: the graph.

$K$: the maximum length of paths we consider.

$I$: the source node from which we want to find all paths to other nodes in the graph.

\Comment{Output}

$D(I, j), j=1\ldots |\mathcal{V}|$: each $D(I, j)$ is a list of path descriptors describing paths between $v_I$ and $v_j$.

\Comment{Variables we use}

$\Phi$: a bit vector of length $|\mathcal{V}|$, where $\Phi(i)$ marks if there is a sensor on node $v_i$.

\Comment {Initialize}
\State $\Phi(I) \gets 1$, $\forall j\neq I, \Phi(j)\gets 0$
\State $\forall j, D(I, j) \gets null$

\Comment {Iteration}

\For{$k=1:K$}
\For {all $i$ in $1:|\mathcal{V}|$ with $\Phi(i)==1$}
\For {all $j$ in $1:|\mathcal{V}|$ with $e_{ij}==1$}
\For {each path $l$ in $D(I,i)$}
\If {$v_j \notin l \; \&\& \; l+e_{ij} \notin D(I,j)$}
\State $D(I,j) = D(I,j)+(l+e_{ij})$
\State $\Phi(j) \gets 1$
\EndIf
\EndFor
\EndFor
\State $\Phi(i) \gets 0$
\EndFor
\EndFor
\State \Return $D(I, j), j=1\ldots |\mathcal{V}|$
\EndProcedure
\end{algorithmic}
\end{algorithm}

\subsection{Optimizing Weights on Paths}
Now that the paths connecting each pair of nodes with similar attributes are found, we continue to optimize the log-likelihood function in Eq.\ref{EqLike} and generate the relationship probabilities $\mathcal{R}$. We derive the gradient for $r_{uv}$ as
\begin{align}
\frac{\partial J}{\partial r^m_{uv}}&=\sum_{i,j}f_m(a_i,a_j)\frac{N_{uv}(v_i\sim v_j)}{p_m(v_{i}\sim v_{j})}.
\label{EqDer}
\end{align}
In the equation, 
\begin{align}
N^m_{uv}(v_i\sim v_j) = \sum_{k=1}^{K}\sum_{h=1}^HI\{l_{kh},e_{uv}\}\frac{p^m_{kh}(v_{i}\sim v_{j})}{r^m_{uv}},
\label{EqN}
\end{align}
where $I\{l_{kh},e_{uv}\}$ is an indicator function computed from path $l_{kh}$. Specifically, $I\{l_{kh},e_{uv}\}$ equals 1 if $l_{kh}$ contains edge $e_{uv}$, and 0 otherwise.
Therefore, $N^m_{uv}(v_i\sim v_j)$ is the sum of the products of all normalized edge weights except for $r^m_{uv}$ along all paths that connect nodes $v_i$ and $v_j$ and also contain edge $e_{uv}$.

With Eq.\ref{EqDer}, we apply standard gradient ascent to solve for $\mathcal{R}$. As can be seen in Eq.\ref{EqN}, for a specific edge $e_{uv}$, the more paths $l_{kh}$'s pass through it  ($I\{l_{kh},e_{uv}\}$ equals 1), the larger the derivative of the corresponding weight $r^m_{uv}$ is, which substantiates our intuition of using paths connecting similarly attributed nodes to profile individual edges. In Eq.\ref{EqR1},  the denominator $d_{h_s}$ exposes an $l$-1 norm on all weights in $\mathcal{R}^m$, encouraging sparse solutions. The penalty arises naturally within the probabilistic model and therefore no heuristic penalty terms to encourage sparsity is necessary.

Consider a specific pair of nodes $v_i$ and $v_j$. Given Eq.\ref{EqR1}, $p_{kh}(v_i\sim v_j)$ is concave in $\mathcal{R}$. Moreover, since different paths connecting $v_i$ and $v_j$ found by our algorithm never share the same edge, $p_{k}(v_i\sim v_j)$ and $p(v_i\sim v_j)$ in Eq.\ref{EqR2} and Eq.\ref{EqR3} are both concave in $\mathcal{R}$. Since log concave is still concave, the log-likelihood function in Eq.\ref{EqR} is a weighted sum of concave functions, which is not globally concave but has an upper bound. However, since $f(a_i,a_j)$'s are usually very sparse in social networks, we find the solution of our algorithm stable and almost always the global optimal in the experiments. 

The runtime of ARP is dominated by finding paths. The runtime of optimization with gradient ascent is linear in $|\mathcal{V}|$. As we study in our experiments, convergence is reached usually within 20 iterations with step size empirically set to 0.05. As discussed before, the time of finding paths is much less than $K|\mathcal{V}|^2$, so the overall time complexity of ARP is $O(K|\mathcal{V}|^2)$, comparable to many advanced attribute profiling and community detection algorithms \cite{chakrabarti2014joint,li2014user,leskovec2012learning}.

\section {Experiments}
\label{sec:exp}
In this section, we evaluate ARP with quantitive experiments and case studies on three real-world datasets. 

\subsection{Experimental Settings}
{\flushleft \bf Datasets.}
The first is the LinkedIn Ego Networks dataset (LEN) from \cite{li2014user}. It includes 268 ego networks, which contain about 19K users and 110K connections. Among them, about 30\% users have attributes of 193 different \textsf{universities} and 375 different \textsf{employers}. 8K connections are labeled by the ego users as carrying the relationships of \textsf{schoolmates}, \textsf{colleagues} or both, based on which we directly perform quantitive performance evaluations. 

The second is the Facebook Ego Networks dataset (FEN) from \cite{leskovec2012learning}. It contains 10 ego networks of about 4K users and 88K connections. We choose all \textsf{hometown}, \textsf{school} and \textsf{employer} attributes out from the total 634 attributes, because they well indicate the relationships of \textsf{townsman}, \textsf{schoolmate} and \textsf{colleague}. Since there are no labeled relationships, we randomly split the users into training and testing sets. We input all users with their connections and attributes in the training set to all compared algorithms, and evaluate the learned relationships on connections between users in the training set and users in the testing set. For quantitive evaluations, we label the relationship of \textsf{townsman} (\textsf{schoolmate}, \textsf{colleague}) on the connections between friends sharing the same attribute value of \textsf{hometown} (\textsf{school}, \textsf{employer}). Thus, there might be multiple relationships on the same connection.

The DBLP data we use were extracted on Jan 1st, 2017, which includes 3.7M publications from 1.8M authors.
We generate nodes as authors and use three publication venues, \textsf{KDD}, \textsf{VLDB} and \textsf{ICML}, as node attributes. A uniform connection is generated between authors who have co-authored at least once in any of the considered venues. Since the authors and attributes are not anonymous, we present insightful case study results on some novel applications to show the effectiveness of our framework.

The attributes captured in both LEN and FEN are incomplete and noisy. In such scenarios, we show that profiling the systematic and complete relationship semantics is generally useful in improving the performance of relationship prediction. 
Although both LEN and FEN are ego-networks, our framework is general to work on any non-ego-networks like DBLP.
Moreover, on DBLP where the attributes are complete and precise, we show that our framework is still advantageous because it provides more insightful results.

{\flushleft \bf Compared algorithms.} The problem of ARP is novel, which is hardly addressed by previous literature. 
As discussed in Sec.~\ref{sec:related}, to comprehensively evaluate ARP, we adapt state-of-the-art algorithms from two groups. 

{\flushleft \it Adapting attribute profiling algorithms.}
These algorithms aim at inferring user attributes based on both known attributes and network structure. Since the relationships they learn are implicit, we need to predict them based on the inferred attributes on the connected users. \xeg, we predict a relationship as \textsf{schoolmate} if the two connected users are inferred with the same \textsf{university} attributes.
\begin{itemize}[leftmargin=15pt]
\setlength\itemsep{0.1em}
\item Relation neighbor classifier (RNC) \cite{macskassy2003simple}: it profiles user attributes \wrt~labeled neighbors without learning.
\item Discriminative relational classifier (DRC) \cite{tang2009relational}: it constructs a modularity-based feature as latent social dimensions to help learn user attributes.
\item EdgeExplain \cite{chakrabarti2014joint}: it improves on traditional label propagation \cite{zhu2002learning,zhu2003semi} by leveraging a softmax function to solve for a global optimal assignment of both user attributes and relationships.
\item BLA \cite{Yang:2017:BJI:3041021.3054181}: it jointly infers link and attribute probabilities by addressing smoothness from two directions on social graphs.
\end{itemize}

{\flushleft \it Adapting community detection algorithms.}
We adapt community detection algorithms that use node attributes to characterize communities, which have a side effect of profiling links at a coarse granularity. We refer to the attribute assignment of each community and predict all relationships based on the most prominent attribute. \xeg, we predict all relationships in a community as \textsf{schoolmate} if a \textsf{university} attribute is the most prominent there.
\begin{itemize}[leftmargin=15pt]
\setlength\itemsep{0.1em}
\item PCL-DC \cite{yang2009combining}: it unifies a conditional model for link and a discriminative model for content analysis. 
\item Circles \cite{leskovec2012learning}: it designs a generative model of edges \wrt~profile similarity to detect overlapping communities.
\item CESNA \cite{yang2013community}: it designs a generative model of edges and attributes to detect overlapping communities.
\item CoProfiling \cite{li2014user}: it profiles attributes and community memberships through iterative coordinate descent.
\end{itemize}

Instead of producing a set of relationship probabilities for each link like ARP, all baselines can only produce categorical labels.

{\flushleft \bf Metrics.}
For performance evaluations, we compute precision, recall and F1 score over all predictions of each relationship as commonly done in related works \cite{li2014user}. The presented results are the averages over 10 times of the same procedures. We also conduct significance tests with $p$-value 0.01.

To further understand the results, we evaluate the relationships profiled by different algorithms \wrt~the systematicness and completeness criteria as we discussed in Sec.~\ref{sec:intro}. We compute the number of all links in the network ($E$), the number of profiled links ($P$) and the number of links profiled with multiple relationships ($M$). To measure the systematicness, we compute $S=P/E$, and to measure completeness, we compute $C=M/P$.

We also measure the actual runtimes of different algorithms on a typical PC with dual 2.3 GHz Intel i7 processors and 8GB memory.

\subsection{Performance Comparison on LEN}
On the LEN dataset, ARP is quantitively evaluated against all baselines. Given a uniform social network, the task is to identify relationships that are discriminatively related to user attributes. Here we aim to identify \textsf{schoolmates}, who are likely to share the same \textsf{university} attributes, and \textsf{colleagues}, who are likely to share the same \textsf{employer} attributes. Evaluation is done on the user labeled relationships. 

We run ARP on the \textsf{university} and \textsf{employer} attributes and predict the probabilities of  \textsf{schoolmate} and \textsf{colleague} relationships on each link. To perform quantitive evaluations, we convert the probabilities into binary predictions for each relationship by thresholding at value $\theta^m$.  
For attribute learning algorithms, we predict \textsf{schoolmates} if the two connected users are inferred with the same \textsf{university} attribute; for community detection algorithms, we predict \textsf{schoolmates} if a certain \textsf{university} is the most prominent attribute for the detected community that contains both connected users. The same is done to predict \textsf{colleagues}.

We select the best parameters for all algorithms via standard 5-fold cross validation. The parameters we set for ARP are $K=3$, $\alpha=0.8$, $\theta^1=0.4$ and $\theta^2=0.7$.
\begin{table}[h]
\small
\centering
 \begin{tabular}{|c|ccc|ccc|}
 \toprule
  \hline
\multirow{2}{*}{Algorithm}&\multicolumn{3}{c|}{Schoolmates}&\multicolumn{3}{c|}{Colleagues}\\
\cline{2-7}
&P&R&F1&P&R&F1\\
\hline
RNC&0.613&0.548&0.579&0.358&0.467&0.405\\
\hline
DRC&0.885&0.472&0.616&0.603&0.442&0.510\\
\hline
EdgeExplain&0.782&0.618&\textit{0.690}&0.530&0.642&\textit{0.581}\\
\hline
BLA&0.648&\textit{0.683}&0.665&0.416&\textit{0.697}&0.521\\
\hline
PCL-DC&0.932&0.498&0.649&0.654&0.516&0.577\\
\hline
Circles&0.937&0.431&0.590&0.512&0.428&0.466\\
\hline
CESNA&0.813&0.492&0.613&0.502&0.538&0.519\\
\hline
CoProfiling&\textbf{0.969}&0.487&0.648&\textit{0.691}&0.453&0.547\\
\hline
\textbf{ARP}&\textit{0.941}&\textbf{0.793}&\textbf{0.861}&\textbf{0.705}&\textbf{0.782}&\textbf{0.742}\\
\hline
  \bottomrule
 \end{tabular}
 \caption{\label{tab:len}\textbf{Performance comparison on LEN.}}
 \vspace{-20pt}
\end{table}

Table \ref{tab:len} shows performance comparison on LEN. The scores all passed the significance tests with p-value 0.01. 
ARP constantly ranks first among the 8 algorithms on F1 score, while other methods have varying performance, which indicates the robustness and universal advantages of our approach on precisely profiling individual links.
By looking into the scores, we find that ARP can effectively improve recall, while maintaining comparable precision to the baselines. 
It shows the effectiveness of our model to systematically and completely profile relationships along paths connecting users with similar attributes. 

We present systematicness and completeness evaluations in Table \ref{table:SC}, where the ratios are averaged through 10 random training-testing splits. As clearly shown, the attribute profiling algorithms usually profile only one relationship for every connection, while the community detection algorithms predict no relationship at all on some connections. ARP is the only one that implements systematic and complete homophily by profiling every connection \wrt~every relationship.

\begin{table}[ht]
\small
\centering
\begin{center}
 \begin{tabular}{|c|cccc|}
 \toprule
  \hline
Algorithm&RNC/DRC&EdgeExplain&BLA&PCL-DL\\
\hline
Systematicness&100\%&100\%&100\%&88.7\%\\
\hline
Completeness&0\%&15.7\%&78.2\%&64.0\%\\
  \hline
  \hline
Algorithm&Circles&CESNA&CoProfiling&ARP\\
\hline
Systematicness&83.6\%&86.2\%&91.4\%&100\%\\
\hline
Completeness&81.2\%&72.9\%&0\%&100\%\\
\hline
  \bottomrule
 \end{tabular}
 \caption{\label{T2}\textbf{Systematicness and completeness of relationships profiled by different algorithms on LEN.}}
  \vspace{-30pt}
\label{table:SC}
  \end{center}
\end{table}

\begin{figure}[th!]
\centering
\includegraphics[width=0.45\textwidth]{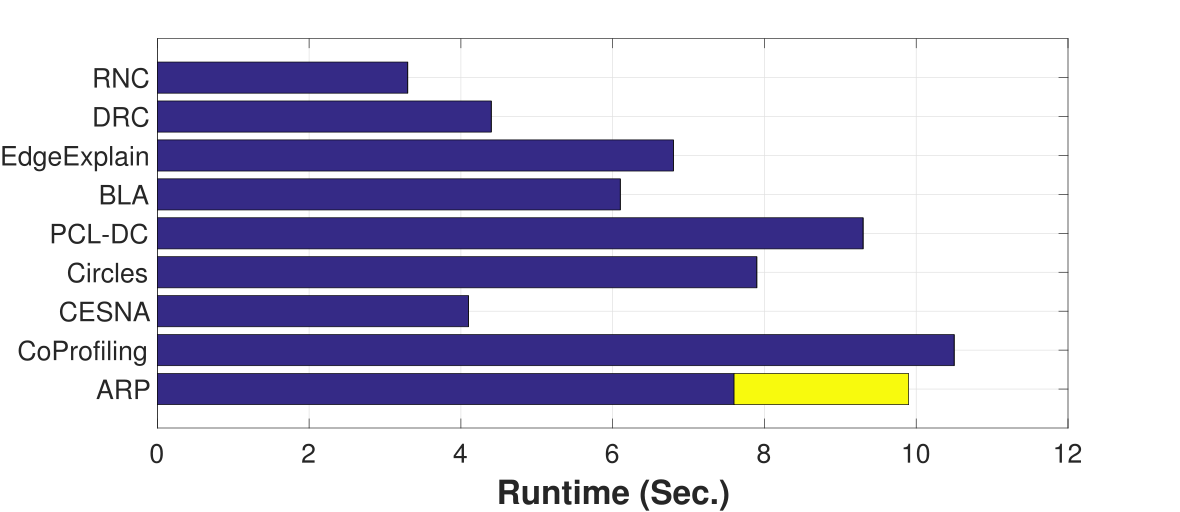}
 \vspace{-10pt}
\caption{\textbf{Runtime comparison on LEN.}}
 \vspace{-10pt}
\label{fig:runtime}
\end{figure}

We present the average runtime of different algorithms on LEN in Figure \ref{fig:runtime}. For ARP, we compare the runtime with and without path caching and reusing, as discussed in Sec.~\ref{sec:algorithm} (the additional runtime without path caching is marked as yellow). The runtime of ARP is comparable to the baselines.

\subsection{Performance and Parameter Study on FEN}
We run experiments on FEN with varying portions of training and testing sets to comprehensively evaluate the performance of ARP. 
We also closely study the impact of the two intrinsic parameters of ARP, \ie, $\alpha$, the decay factor, and $K$, the maximum length of paths we consider. 

To compute the F1 scores, the similar process for LEN has been done to all compared algorithms to yield a binary prediction for each of the \textsf{townsman}, \textsf{schoolmate} and \textsf{colleague} relationships on each connection.

\begin{figure}[h!]
\centering
\subfigure[Varying $\alpha$ while $K=3$]{
\includegraphics[width=0.23\textwidth]{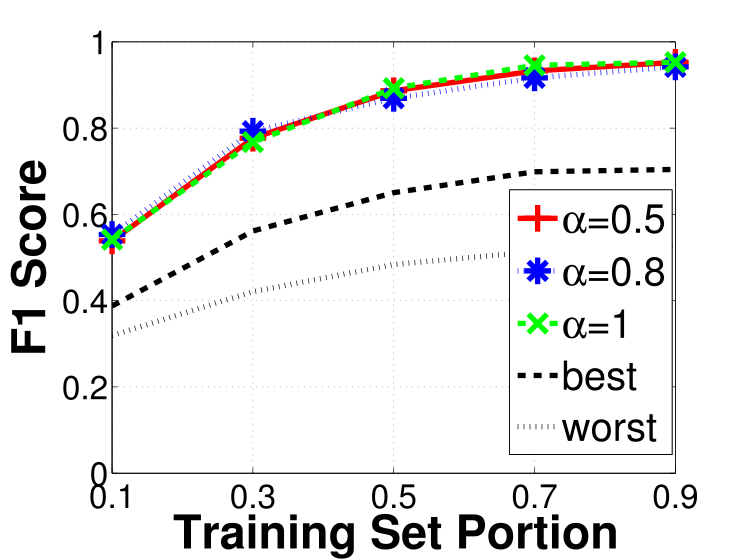}}
\subfigure[Varying $K$ while $\alpha=0.8$]{
\includegraphics[width=0.23\textwidth]{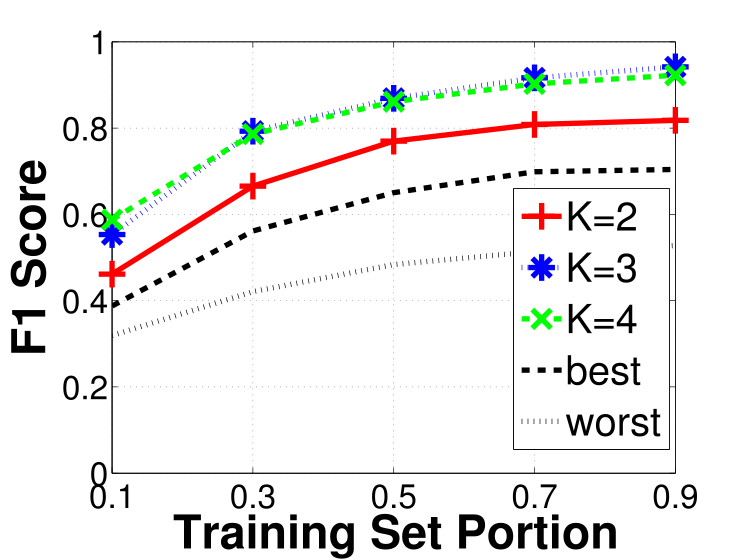}}
 \vspace{-10pt}
\caption{\textbf{Performance study with baselines on FEN.}}
 \vspace{-10pt}
\label{fig:parameter}
\end{figure}

Figure \ref{fig:parameter} shows experimental results on FEN for \textsf{townsman} with $\theta^1=0.2$. The results for \textsf{schoolmates} and \textsf{colleagues} are similar. In Figure \ref{fig:parameter}(a), the decay factor $\alpha$ does not significantly influence the performances. This is probably because we only consider short paths. In Figure \ref{fig:parameter}(b), when $K$ is set to 2, only two-step paths are considered, which leads to quite poor results. When $K$ is set to larger values like 3 and 4, the holistic modeling approach becomes effective and the results are much better. Note that $K=3$ and $K=4$ always yield similar results, which indicates that the importance of paths is dominated by short ones. By setting $K$ to small values like 3, we can run ARP efficiently by avoiding irrelevant edges.

\subsection{Case Study on DBLP}
One advantage of ARP over the compared algorithms is that it can estimate the probability of each connection to carry each relationship. On LEN and FEN, we convert the probabilities into binary outputs in order to present quantitive comparisons with the baselines. However, the application of ARP is much broader than binary predictions. We use DBLP to present some insightful results derived from the relationship probabilities, which only ARP can generate.

Consider some interesting novel applications on DBLP. One of them is to find out people's closest co-authors within different research fields. \xeg, two authors might study similar problems in data mining, but very different problems in database. Thus, how can we identify people's closest co-authors given a specific field? Another interesting application is to identify the closest pairs of authors within each field of study, \ie, who study the closest problems and collaborate most in a specific field? Considering specific relationships, such problems are novel and naturally different from general graph ranking.

We show that problems like these are direct applications of ARP. By modeling publication venues as user attributes and co-authorship as user connections, ARP accurately computes the closeness among authors \wrt~different fields.

Consider three representative venues that correspond to three different but related fields. 
Table 3 shows the relationship specific closeness learned by ARP and normalized into multinomial distributions over each pair of authors. While relationships can be multiple and vary across connections, ARP completely retrieves them in all aspects. 

\begin{table}[h]
 \small
 \centering
 \begin{tabular}{|c|ccc|}
 \toprule
  \hline
\textbf{Authors}&\textbf{KDD}&\textbf{VLDB}&\textbf{ICML}\\
\hline
Jiawei Han, Philip S. Yu&\textbf{0.65}&\textit{0.35}&0\\
Jiawei Han, Xiaolei Li&\textit{0.04}&\textbf{0.96}&0\\
Jiawei Han, Tianbao Yang&\textit{0.17}&0&\textbf{0.83}\\
Christos Faloutsos, Hanghang Tong&\textbf{0.86}&\textit{0.13}&0.01\\
Divesh Srivastava, H. V. Jagadish&\textit{0.03}&\textbf{0.97}&0\\
Corinna Cortes, Mehryar Mohri&\textit{0.14}&0&\textbf{0.86}\\
\hline
  \bottomrule
 \end{tabular}
 \caption{\label{T2}\textbf{Multi-aspect relationships among authors.}}
  \vspace{-10pt}
\end{table}

\begin{table}[h]
 \small
 \centering
 \begin{tabular}{|ccc|}
 \toprule
  \hline
\textbf{Authors} & \textbf{(A) Holistic} & \textbf{(B) Single}\\
\hline
Jiawei Han, Chi Wang&1.00/1st&1.00/1st\\
Christos Faloutsos, Hanghang Tong&0.95/2nd&1.00/1st\\
Hynne Hsu, Mong-Li Lee&0.93/3rd&0.72/10th\\
Jiawei Han, Philip S. Yu&0.90/4th&0.63/18th\\
Christos Faloutsos, Lei Li&0.85/5th&0.72/10th\\
\hline
  \bottomrule
 \end{tabular}
 \caption{\label{T2}\textbf{Rank list of closest authors on KDD.}}
  \vspace{-20pt}
\end{table}

In Table 4, pairs of authors are ranked with their relative closeness in the research field of data mining \wrt~the KDD conference. Column (A) shows the results of holistic modeling, where we set $K=3$ and $\alpha=0.8$ to consider indirect collaborations. The results are intuitive because the top ranked pairs of authors are indeed those who collaborate most in the field. To show that the results in Column (A) are non-trivial as cannot be simply computed by counting the number of collaborated papers, we also provide in Column (B) the results without holistic modeling, which are less intuitive. \xeg, although Han and Yu work quite closely on data mining, the closeness between them decreases from 0.90 to 0.63 and their rank drops from $4th$ to $18th$, merely because their many indirect collaborations are ignored. The situations are similar for many other pairs such as Faloutsos and Li.

Due to space limit, please refer to our anonymous Github project\footnote{https://github.com/yangji9181/ARP2017} to explore more interesting novel applications and visualizations enabled by ARP. The codes are also available under the same directory.


\section{Conclusion}
While ARP is a novel problem that can be viewed as an essential part of problems such as attribute learning and community detection, we emphasize that this problem itself is important, complex and of great research value.
As a unique solution, we propose to learn relationship semantics in a principled probabilistic way, which characterizes the formation of each user connection in social networks based on user attributes. 
Since ARP enables automatic labeling of relationships in an unsupervised way, the roles that different relationships play in various networks can be rigorously studied, such as promoting certain messages and shaping specific groups.


%
	 
\bibliographystyle{ACM-Reference-Format}
\bibliography{arp} 


\begin{thebibliography}{00}


\ifx \showCODEN    \undefined \def \showCODEN     #1{\unskip}     \fi
\ifx \showDOI      \undefined \def \showDOI       #1{{\tt DOI:}\penalty0{#1}\ }
  \fi
\ifx \showISBNx    \undefined \def \showISBNx     #1{\unskip}     \fi
\ifx \showISBNxiii \undefined \def \showISBNxiii  #1{\unskip}     \fi
\ifx \showISSN     \undefined \def \showISSN      #1{\unskip}     \fi
\ifx \showLCCN     \undefined \def \showLCCN      #1{\unskip}     \fi
\ifx \shownote     \undefined \def \shownote      #1{#1}          \fi
\ifx \showarticletitle \undefined \def \showarticletitle #1{#1}   \fi
\ifx \showURL      \undefined \def \showURL       #1{#1}          \fi
\providecommand\bibfield[2]{#2}
\providecommand\bibinfo[2]{#2}
\providecommand\natexlab[1]{#1}
\providecommand\showeprint[2][]{arXiv:#2}

\bibitem[\protect\citeauthoryear{Backstrom and Leskovec}{Backstrom and
  Leskovec}{2011}]%
        {backstrom2011supervised}
\bibfield{author}{\bibinfo{person}{Lars Backstrom} {and} \bibinfo{person}{Jure
  Leskovec}.} \bibinfo{year}{2011}\natexlab{}.
\newblock \showarticletitle{Supervised random walks: predicting and
  recommending links in social networks}. In \bibinfo{booktitle}{{\em WSDM}}.
  \bibinfo{pages}{635--644}.
\newblock


\bibitem[\protect\citeauthoryear{Chakrabarti, Funiak, Chang, and
  Macskassy}{Chakrabarti et~al\mbox{.}}{2014}]%
        {chakrabarti2014joint}
\bibfield{author}{\bibinfo{person}{Deepayan Chakrabarti},
  \bibinfo{person}{Stanislav Funiak}, \bibinfo{person}{Jonathan Chang}, {and}
  \bibinfo{person}{Sofus Macskassy}.} \bibinfo{year}{2014}\natexlab{}.
\newblock \showarticletitle{Joint Inference of Multiple Label Types in Large
  Networks}. In \bibinfo{booktitle}{{\em ICML}}. \bibinfo{pages}{874--882}.
\newblock


\bibitem[\protect\citeauthoryear{Choroma{\'n}ski, Matuszak, and
  Mi?kisz}{Choroma{\'n}ski et~al\mbox{.}}{2013}]%
        {choromanski2013scale}
\bibfield{author}{\bibinfo{person}{Krzysztof Choroma{\'n}ski},
  \bibinfo{person}{Micha{\l} Matuszak}, {and} \bibinfo{person}{Jacek Mi?kisz}.}
  \bibinfo{year}{2013}\natexlab{}.
\newblock \showarticletitle{Scale-free graph with preferential attachment and
  evolving internal vertex structure}.
\newblock \bibinfo{journal}{{\em Journal of Statistical Physics\/}}
  \bibinfo{volume}{151}, \bibinfo{number}{6} (\bibinfo{year}{2013}),
  \bibinfo{pages}{1175--1183}.
\newblock


\bibitem[\protect\citeauthoryear{Fang, Chang, and Lauw}{Fang
  et~al\mbox{.}}{2014}]%
        {fang2014graph}
\bibfield{author}{\bibinfo{person}{Yuan Fang}, \bibinfo{person}{Kevin
  Chen-Chuan Chang}, {and} \bibinfo{person}{Hady~Wirawan Lauw}.}
  \bibinfo{year}{2014}\natexlab{}.
\newblock \showarticletitle{Graph-based Semi-supervised Learning: Realizing
  Pointwise Smoothness Probabilistically}. In \bibinfo{booktitle}{{\em ICML}}.
\newblock


\bibitem[\protect\citeauthoryear{Han and Tang}{Han and Tang}{2015}]%
        {han2015probabilistic}
\bibfield{author}{\bibinfo{person}{Yu Han} {and} \bibinfo{person}{Jie Tang}.}
  \bibinfo{year}{2015}\natexlab{}.
\newblock \showarticletitle{Probabilistic community and role model for social
  networks}. In \bibinfo{booktitle}{{\em Proceedings of the 21th ACM SIGKDD
  International Conference on Knowledge Discovery and Data Mining}}. ACM,
  \bibinfo{pages}{407--416}.
\newblock


\bibitem[\protect\citeauthoryear{He, Carbonell, and Liu}{He
  et~al\mbox{.}}{2007}]%
        {he2007graph}
\bibfield{author}{\bibinfo{person}{Jingrui He}, \bibinfo{person}{Jaime~G
  Carbonell}, {and} \bibinfo{person}{Yan Liu}.}
  \bibinfo{year}{2007}\natexlab{}.
\newblock \showarticletitle{Graph-Based Semi-Supervised Learning as a
  Generative Model.}. In \bibinfo{booktitle}{{\em IJCAI}},
  Vol.~\bibinfo{volume}{7}. \bibinfo{pages}{2492--2497}.
\newblock


\bibitem[\protect\citeauthoryear{Li, Wang, and Chang}{Li et~al\mbox{.}}{2014}]%
        {li2014user}
\bibfield{author}{\bibinfo{person}{Rui Li}, \bibinfo{person}{Chi Wang}, {and}
  \bibinfo{person}{Kevin Chen-Chuan Chang}.} \bibinfo{year}{2014}\natexlab{}.
\newblock \showarticletitle{User profiling in an ego network: co-profiling
  attributes and relationships}. In \bibinfo{booktitle}{{\em WWW}}.
  \bibinfo{pages}{819--830}.
\newblock


\bibitem[\protect\citeauthoryear{Lin, Yang, He, and Ye}{Lin
  et~al\mbox{.}}{2014}]%
        {lin2014geodesic}
\bibfield{author}{\bibinfo{person}{Binbin Lin}, \bibinfo{person}{Ji Yang},
  \bibinfo{person}{Xiaofei He}, {and} \bibinfo{person}{Jieping Ye}.}
  \bibinfo{year}{2014}\natexlab{}.
\newblock \showarticletitle{Geodesic Distance Function Learning via Heat Flow
  on Vector Fields}. In \bibinfo{booktitle}{{\em ICML}}.
\newblock


\bibitem[\protect\citeauthoryear{Lov{\'a}sz}{Lov{\'a}sz}{1993}]%
        {lovasz1993random}
\bibfield{author}{\bibinfo{person}{L{\'a}szl{\'o} Lov{\'a}sz}.}
  \bibinfo{year}{1993}\natexlab{}.
\newblock \showarticletitle{Random walks on graphs: A survey}.
\newblock \bibinfo{journal}{{\em Combinatorics, Paul erdos is eighty\/}}
  \bibinfo{volume}{2}, \bibinfo{number}{1} (\bibinfo{year}{1993}),
  \bibinfo{pages}{1--46}.
\newblock


\bibitem[\protect\citeauthoryear{Macskassy and Provost}{Macskassy and
  Provost}{2003}]%
        {macskassy2003simple}
\bibfield{author}{\bibinfo{person}{Sofus~A Macskassy} {and}
  \bibinfo{person}{Foster Provost}.} \bibinfo{year}{2003}\natexlab{}.
\newblock \bibinfo{booktitle}{{\em A simple relational classifier}}.
\newblock \bibinfo{type}{{T}echnical {R}eport}. \bibinfo{institution}{DTIC
  Document}.
\newblock


\bibitem[\protect\citeauthoryear{Mcauley and Leskovec}{Mcauley and
  Leskovec}{2012}]%
        {leskovec2012learning}
\bibfield{author}{\bibinfo{person}{Julian~J Mcauley} {and}
  \bibinfo{person}{Jure Leskovec}.} \bibinfo{year}{2012}\natexlab{}.
\newblock \showarticletitle{Learning to discover social circles in ego
  networks}. In \bibinfo{booktitle}{{\em NIPS}}. \bibinfo{pages}{539--547}.
\newblock


\bibitem[\protect\citeauthoryear{McPherson, Smith-Lovin, and Cook}{McPherson
  et~al\mbox{.}}{2001}]%
        {mcpherson2001birds}
\bibfield{author}{\bibinfo{person}{Miller McPherson}, \bibinfo{person}{Lynn
  Smith-Lovin}, {and} \bibinfo{person}{James~M Cook}.}
  \bibinfo{year}{2001}\natexlab{}.
\newblock \showarticletitle{Birds of a feather: Homophily in social networks}.
\newblock \bibinfo{journal}{{\em Annual review of sociology\/}}
  (\bibinfo{year}{2001}), \bibinfo{pages}{415--444}.
\newblock


\bibitem[\protect\citeauthoryear{Mislove, Viswanath, Gummadi, and
  Druschel}{Mislove et~al\mbox{.}}{2010}]%
        {mislove2010you}
\bibfield{author}{\bibinfo{person}{Alan Mislove}, \bibinfo{person}{Bimal
  Viswanath}, \bibinfo{person}{Krishna~P Gummadi}, {and} \bibinfo{person}{Peter
  Druschel}.} \bibinfo{year}{2010}\natexlab{}.
\newblock \showarticletitle{You are who you know: inferring user profiles in
  online social networks}. In \bibinfo{booktitle}{{\em WSDM}}.
  \bibinfo{pages}{251--260}.
\newblock


\bibitem[\protect\citeauthoryear{Page, Brin, Motwani, and Winograd}{Page
  et~al\mbox{.}}{1999}]%
        {page1999pagerank}
\bibfield{author}{\bibinfo{person}{Lawrence Page}, \bibinfo{person}{Sergey
  Brin}, \bibinfo{person}{Rajeev Motwani}, {and} \bibinfo{person}{Terry
  Winograd}.} \bibinfo{year}{1999}\natexlab{}.
\newblock \showarticletitle{The PageRank citation ranking: Bringing order to
  the web.}
\newblock  (\bibinfo{year}{1999}).
\newblock


\bibitem[\protect\citeauthoryear{Rakesh, Lee, and Reddy}{Rakesh
  et~al\mbox{.}}{2016}]%
        {rakesh2016probabilistic}
\bibfield{author}{\bibinfo{person}{Vineeth Rakesh}, \bibinfo{person}{Wang-Chien
  Lee}, {and} \bibinfo{person}{Chandan~K Reddy}.}
  \bibinfo{year}{2016}\natexlab{}.
\newblock \showarticletitle{Probabilistic Group Recommendation Model for
  Crowdfunding Domains}. In \bibinfo{booktitle}{{\em Proceedings of the Ninth
  ACM International Conference on Web Search and Data Mining}}. ACM,
  \bibinfo{pages}{257--266}.
\newblock


\bibitem[\protect\citeauthoryear{Ruan, Fuhry, and Parthasarathy}{Ruan
  et~al\mbox{.}}{2013}]%
        {ruan2013efficient}
\bibfield{author}{\bibinfo{person}{Yiye Ruan}, \bibinfo{person}{David Fuhry},
  {and} \bibinfo{person}{Srinivasan Parthasarathy}.}
  \bibinfo{year}{2013}\natexlab{}.
\newblock \showarticletitle{Efficient community detection in large networks
  using content and links}. In \bibinfo{booktitle}{{\em WWW}}.
  \bibinfo{pages}{1089--1098}.
\newblock


\bibitem[\protect\citeauthoryear{Rubin}{Rubin}{1978}]%
        {rubin1978enumerating}
\bibfield{author}{\bibinfo{person}{Frank Rubin}.}
  \bibinfo{year}{1978}\natexlab{}.
\newblock \showarticletitle{Enumerating all simple paths in a graph}.
\newblock \bibinfo{journal}{{\em ITCS\/}} \bibinfo{volume}{25},
  \bibinfo{number}{8} (\bibinfo{year}{1978}), \bibinfo{pages}{641--642}.
\newblock


\bibitem[\protect\citeauthoryear{Sachan, Dubey, Srivastava, Xing, and
  Hovy}{Sachan et~al\mbox{.}}{2014}]%
        {sachan2014spatial}
\bibfield{author}{\bibinfo{person}{Mrinmaya Sachan}, \bibinfo{person}{Avinava
  Dubey}, \bibinfo{person}{Shashank Srivastava}, \bibinfo{person}{Eric~P Xing},
  {and} \bibinfo{person}{Eduard Hovy}.} \bibinfo{year}{2014}\natexlab{}.
\newblock \showarticletitle{Spatial compactness meets topical consistency:
  jointly modeling links and content for community detection}. In
  \bibinfo{booktitle}{{\em Proceedings of the 7th ACM international conference
  on Web search and data mining}}. ACM, \bibinfo{pages}{503--512}.
\newblock


\bibitem[\protect\citeauthoryear{{\c{S}}im{\c{s}}ek and
  Jensen}{{\c{S}}im{\c{s}}ek and Jensen}{2008}]%
        {csimcsek2008navigating}
\bibfield{author}{\bibinfo{person}{{\"O}zg{\"u}r {\c{S}}im{\c{s}}ek} {and}
  \bibinfo{person}{David Jensen}.} \bibinfo{year}{2008}\natexlab{}.
\newblock \showarticletitle{Navigating networks by using homophily and degree}.
\newblock \bibinfo{journal}{{\em Proceedings of the National Academy of
  Sciences\/}} \bibinfo{volume}{105}, \bibinfo{number}{35}
  (\bibinfo{year}{2008}), \bibinfo{pages}{12758--12762}.
\newblock


\bibitem[\protect\citeauthoryear{Subramanya and Bilmes}{Subramanya and
  Bilmes}{2011}]%
        {subramanya2011semi}
\bibfield{author}{\bibinfo{person}{Amarnag Subramanya} {and}
  \bibinfo{person}{Jeff Bilmes}.} \bibinfo{year}{2011}\natexlab{}.
\newblock \showarticletitle{Semi-supervised learning with measure propagation}.
\newblock \bibinfo{journal}{{\em Journal of Machine Learning Research\/}}
  \bibinfo{volume}{12}, \bibinfo{number}{Nov} (\bibinfo{year}{2011}),
  \bibinfo{pages}{3311--3370}.
\newblock


\bibitem[\protect\citeauthoryear{Tang and Liu}{Tang and Liu}{2009}]%
        {tang2009relational}
\bibfield{author}{\bibinfo{person}{Lei Tang} {and} \bibinfo{person}{Huan Liu}.}
  \bibinfo{year}{2009}\natexlab{}.
\newblock \showarticletitle{Relational learning via latent social dimensions}.
  In \bibinfo{booktitle}{{\em Proceedings of the 15th ACM SIGKDD international
  conference on Knowledge discovery and data mining}}. ACM,
  \bibinfo{pages}{817--826}.
\newblock


\bibitem[\protect\citeauthoryear{Watts and Strogatz}{Watts and
  Strogatz}{1998}]%
        {watts1998collective}
\bibfield{author}{\bibinfo{person}{Duncan~J Watts} {and}
  \bibinfo{person}{Steven~H Strogatz}.} \bibinfo{year}{1998}\natexlab{}.
\newblock \showarticletitle{Collective dynamics of small-world networks}.
\newblock \bibinfo{journal}{{\em nature\/}} \bibinfo{volume}{393},
  \bibinfo{number}{6684} (\bibinfo{year}{1998}), \bibinfo{pages}{440--442}.
\newblock


\bibitem[\protect\citeauthoryear{Yang, Zhong, Li, and Jie}{Yang
  et~al\mbox{.}}{2017}]%
        {Yang:2017:BJI:3041021.3054181}
\bibfield{author}{\bibinfo{person}{Carl Yang}, \bibinfo{person}{Lin Zhong},
  \bibinfo{person}{Li-Jia Li}, {and} \bibinfo{person}{Luo Jie}.}
  \bibinfo{year}{2017}\natexlab{}.
\newblock \showarticletitle{Bi-directional Joint Inference for User Links and
  Attributes on Large Social Graphs}. In \bibinfo{booktitle}{{\em Proceedings
  of the 26th International Conference on World Wide Web Companion}}.
  \bibinfo{pages}{564--573}.
\newblock


\bibitem[\protect\citeauthoryear{Yang, McAuley, and Leskovec}{Yang
  et~al\mbox{.}}{2013}]%
        {yang2013community}
\bibfield{author}{\bibinfo{person}{Jaewon Yang}, \bibinfo{person}{Julian
  McAuley}, {and} \bibinfo{person}{Jure Leskovec}.}
  \bibinfo{year}{2013}\natexlab{}.
\newblock \showarticletitle{Community detection in networks with node
  attributes}. In \bibinfo{booktitle}{{\em ICDM}}. \bibinfo{pages}{1151--1156}.
\newblock


\bibitem[\protect\citeauthoryear{Yang, Long, Smola, Sadagopan, Zheng, and
  Zha}{Yang et~al\mbox{.}}{2011}]%
        {yang2011like}
\bibfield{author}{\bibinfo{person}{Shuang-Hong Yang}, \bibinfo{person}{Bo
  Long}, \bibinfo{person}{Alex Smola}, \bibinfo{person}{Narayanan Sadagopan},
  \bibinfo{person}{Zhaohui Zheng}, {and} \bibinfo{person}{Hongyuan Zha}.}
  \bibinfo{year}{2011}\natexlab{}.
\newblock \showarticletitle{Like like alike: joint friendship and interest
  propagation in social networks}. In \bibinfo{booktitle}{{\em WWW}}.
  \bibinfo{pages}{537--546}.
\newblock


\bibitem[\protect\citeauthoryear{Yang, Jin, Chi, and Zhu}{Yang
  et~al\mbox{.}}{2009}]%
        {yang2009combining}
\bibfield{author}{\bibinfo{person}{Tianbao Yang}, \bibinfo{person}{Rong Jin},
  \bibinfo{person}{Yun Chi}, {and} \bibinfo{person}{Shenghuo Zhu}.}
  \bibinfo{year}{2009}\natexlab{}.
\newblock \showarticletitle{Combining link and content for community detection:
  a discriminative approach}. In \bibinfo{booktitle}{{\em SIGKDD}}.
  \bibinfo{pages}{927--936}.
\newblock


\bibitem[\protect\citeauthoryear{Yu, Ren, Sun, Sturt, Khandelwal, Gu, Norick,
  and Han}{Yu et~al\mbox{.}}{2013}]%
        {yu2013recommendation}
\bibfield{author}{\bibinfo{person}{Xiao Yu}, \bibinfo{person}{Xiang Ren},
  \bibinfo{person}{Yizhou Sun}, \bibinfo{person}{Bradley Sturt},
  \bibinfo{person}{Urvashi Khandelwal}, \bibinfo{person}{Quanquan Gu},
  \bibinfo{person}{Brandon Norick}, {and} \bibinfo{person}{Jiawei Han}.}
  \bibinfo{year}{2013}\natexlab{}.
\newblock \showarticletitle{Recommendation in heterogeneous information
  networks with implicit user feedback}. In \bibinfo{booktitle}{{\em
  Proceedings of the 7th ACM conference on Recommender systems}}. ACM,
  \bibinfo{pages}{347--350}.
\newblock


\bibitem[\protect\citeauthoryear{Zhou, Bousquet, Lal, Weston, and
  Sch{\"o}lkopf}{Zhou et~al\mbox{.}}{2004}]%
        {zhou2004learning}
\bibfield{author}{\bibinfo{person}{Dengyong Zhou}, \bibinfo{person}{Olivier
  Bousquet}, \bibinfo{person}{Thomas~Navin Lal}, \bibinfo{person}{Jason
  Weston}, {and} \bibinfo{person}{Bernhard Sch{\"o}lkopf}.}
  \bibinfo{year}{2004}\natexlab{}.
\newblock \showarticletitle{Learning with local and global consistency}.
\newblock \bibinfo{journal}{{\em NIPS\/}} \bibinfo{volume}{16},
  \bibinfo{number}{16} (\bibinfo{year}{2004}), \bibinfo{pages}{321--328}.
\newblock


\bibitem[\protect\citeauthoryear{Zhu, Fang, Chang, and Ying}{Zhu
  et~al\mbox{.}}{2013}]%
        {zhu2013incremental}
\bibfield{author}{\bibinfo{person}{Fanwei Zhu}, \bibinfo{person}{Yuan Fang},
  \bibinfo{person}{Kevin Chen-Chuan Chang}, {and} \bibinfo{person}{Jing Ying}.}
  \bibinfo{year}{2013}\natexlab{}.
\newblock \showarticletitle{Incremental and accuracy-aware personalized
  pagerank through scheduled approximation}.
\newblock \bibinfo{journal}{{\em VLDB\/}} \bibinfo{volume}{6},
  \bibinfo{number}{6} (\bibinfo{year}{2013}), \bibinfo{pages}{481--492}.
\newblock


\bibitem[\protect\citeauthoryear{Zhu and Ghahramani}{Zhu and
  Ghahramani}{2002}]%
        {zhu2002learning}
\bibfield{author}{\bibinfo{person}{Xiaojin Zhu} {and} \bibinfo{person}{Zoubin
  Ghahramani}.} \bibinfo{year}{2002}\natexlab{}.
\newblock \bibinfo{booktitle}{{\em Learning from labeled and unlabeled data
  with label propagation}}.
\newblock \bibinfo{type}{{T}echnical {R}eport}.
  \bibinfo{institution}{Citeseer}.
\newblock


\bibitem[\protect\citeauthoryear{Zhu, Ghahramani, Lafferty, et~al\mbox{.}}{Zhu
  et~al\mbox{.}}{2003}]%
        {zhu2003semi}
\bibfield{author}{\bibinfo{person}{Xiaojin Zhu}, \bibinfo{person}{Zoubin
  Ghahramani}, \bibinfo{person}{John Lafferty}, {and}
  \bibinfo{person}{others}.} \bibinfo{year}{2003}\natexlab{}.
\newblock \showarticletitle{Semi-supervised learning using gaussian fields and
  harmonic functions}. In \bibinfo{booktitle}{{\em ICML}},
  Vol.~\bibinfo{volume}{3}. \bibinfo{pages}{912--919}.
\newblock


\end{thebibliography}

\end{document}